\documentclass[pre,twocolumn,showpacs,preprintnumbers,amsmath,amssymb]{revtex4}
\usepackage{graphics}

\begin{document}

\title{ Accurate structure factors from pseudopotential methods }
\author{J R Trail}\email{j.r.trail@bath.ac.uk}
\author{D M Bird}
\affiliation{Department of Physics, University of Bath, Bath BA2 7AY, UK}

\date{June, 1999}

\begin{abstract}
Highly accurate experimental structure factors of silicon are
available in the literature, and these provide the ideal test for any
\emph{ab initio} method for the construction of the all-electron
charge density.  In a recent paper
[J. R. Trail and D. M. Bird, Phys.\ Rev.\ B {\bf 60}, 7863 (1999)]
a method has been developed for obtaining an accurate
all-electron charge density from a first principles pseudopotential
calculation by reconstructing the core region of an atom of choice.
Here this method is applied to bulk silicon, and structure factors are
derived and compared with experimental and Full-potential Linear
Augmented Plane Wave results (FLAPW).  We also compare with the result
of assuming the core region is spherically symmetric, and with the
result of constructing a charge density from the pseudo-valence
density + frozen core electrons.  Neither of these approximations
provide accurate charge densities.  The aspherical reconstruction is
found to be as accurate as FLAPW results, and reproduces the residual
error between the FLAPW and experimental results.
\end{abstract}

 
\maketitle

\section{Introduction}
Pseudopotential methods, particularly within the framework of total
energy plane-wave calculations, are extremely powerful for the
\emph{ab initio} description of large system of atoms due to their
computational efficiency and suitability for structural
optimisation\cite{payne92}.  However, they do not yield the correct
charge density of the system studied, but a `pseudo' charge density
that does not include core electrons and is incorrect close to atomic
nuclei.  This precludes the direct application of these methods to the
prediction of any properties of the material that depend directly on
the charge density, such as hyperfine couplings or X-ray structure
factors.  In this paper we reconstruct the all-electron charge density
for bulk silicon from a pseudopotential calculation using the method
described by Trail and Bird\cite{trail99} (hereafter referred to as
I), and from this we derive the X-ray structure factors.  These are
then compared with experimental results and the results of other
theoretical approximations and methods to assess the importance of
various assumptions often made in the calculation of structure
factors, and to evaluate the success of the reconstruction method.

Previous methods \cite{gardner86,vackar94,kuzmiak91,meyer95} for
solving this reconstruction problem have relied on the assumption that
the potential in the core region is spherically symmetric in order to
decouple the differential equations that must be solved, and in many
cases the charge density itself is assumed to be spherical.  The
method used here does not require this to be the case, and we compare
the structure factors resulting from assuming spherical symmetry to
justify the extra effort necessary to develop an aspherical
reconstruction procedure.  The reconstruction method itself is based
around the embedding approach of Inglesfield \cite{inglesfield81}.
Results from this localised calculation are used to replace the
pseudo-charge density where this is incorrect, leading to the required
structure factors.

In section \ref{sec:method} a brief summary of the reconstruction
approach is given - a full description can be found in I.  We describe
how we obtain the structure factors from the reconstruction in section
\ref{sec:factors}, and compare the reconstruction results with
accurate experimental and theoretical structure factors.  Rydberg
atomic units are used throughout the paper.

\section{Reconstruction Method}
\label{sec:method}
The first step in the reconstruction procedure is to obtain an
accurate approximation for the real space single particle Green
function of the substrate system, in this case bulk silicon.  We begin
with a total energy pseudopotential calculation performed with a
plane-wave basis and using the Local Density Approximation (LDA) for
exchange and correlation.  A plane-wave energy cut-off of $400$ eV is
used and $28$ Monkhorst-Pack $\mathbf k$-points\cite{monkhorst76} are
included in the irreducible wedge of the FCC Brillouin zone.  These
values are more than sufficient to obtain essentially perfect
convergence of the self-consistent density and potential, which allows
us to attribute any errors in our results to the reconstruction
procedure.  A norm-conserving Kerker\cite{kerker80} pseudopotential is
used, with a maximum core radius of $2.0$ au.  As explained in I the
method requires $r_c$ to be less than half the nearest-neighbour
atomic separation in the crystal.  The resulting self-consistent
potential is used to obtain a set of eigenstates by direct matrix
diagonalisation, at $240\ \mathbf k$-points in the irreducible wedge
of the Brillouin zone and with a $200$ eV plane-wave energy cut-off.
Careful tests have been carried out to confirm that these values are
sufficient to provide structure factors with a precision of order
$\sim 1$ milli-electrons/atom.  The spectral representation
\cite{morse53} is used to construct a Green function from the set of
plane-wave states.  This Green function is then used to obtain an
embedding potential, which is a term that is added to the Kohn-Sham
Hamiltonian for the localised core region of an atom of interest.  The
effect of the embedding potential is to take into account the lattice
of pseudo-atoms surrounding the chosen atom.  The localised embedded
Hamiltonian is then solved self-consistently (again using the LDA) to
obtain the Green function of the embedded system, from which the
charge density in the core region can be obtained (see paper I).

The reconstruction is performed using the same parameters as in I,
with the embedding radius chosen as the `touching spheres' radius (in
this case $r_s=2.222$ au).  Again, convergence with respect to all
parameters has been thoroughly checked.  The final result we arrive at
is for the charge density of a single all-electron atom embedded in a
lattice of pseudo-atoms.  This information can be used together with
the original pseudo-charge density to construct an accurate
all-electron charge density and hence the structure factors for the
crystal.

\section{Structure Factors}
\label{sec:factors}
Extremely accurate experimental structure factors for silicon have
been available in the literature for some time
\cite{cumming88,aldred73,teworte84,saka86}.  These results have been
used by a number of workers to assess the accuracy of parameterised
models \cite{deutsch92}, FLAPW and other \emph{ab initio} methods
\cite{lu93} and Generalised Gradient Approximations to the
exchange-correlation potential \cite{zuo97}.  In view of the accuracy
and range of data available, both experimental and theoretical, the
reconstructed silicon charge densities are here used to construct
structure factors for comparison with experimental data and the
results of FLAPW calculations.

\subsection{Structure Factors from Reconstructed Charge Densities}
To obtain the structure factors we require the Fourier coefficients of
the charge density,
\begin{equation}
\rho_{total}({\mathbf g})=\frac{1}{\Omega}
\int_{\Omega}\rho_{total}({\mathbf r})e^{-i{\mathbf g}.{\mathbf
r}}d^3{\mathbf r}
\label{eq1}
\end{equation}
where $\Omega$ denotes the unit cell volume and $\rho_{total}({\mathbf
r})$ is the total real space charge density.  $\rho_{total}$ consists
of the original pseudo-charge density between atoms, and the
reconstructed total charge density within the embedding sphere
surrounding each atom.  Since this integral is a linear operation on
the charge density it is possible to subtract the contribution to the
pseudo-density from the embedding regions around each atom and add on
the contributions from a reconstruction calculation.  This gives the
expression
\begin{equation}
\sum_{i}\left[ \alpha^{recon}_{{\mathbf s}_i}({\mathbf
g})-\alpha^{pseudo}_{{\mathbf s}_i}({\mathbf g}) \right]
\label{eq2}
\end{equation}
where ${\mathbf s}_i$ are the position vectors of the atoms in the
unit cell.  The quantities $\alpha^{recon}$ and $\alpha^{pseudo}$ are
the Fourier integrals of the reconstructed and pseudo-densities
respectively, carried out over the reconstruction sphere surrounding
each atom, and are given by
\begin{equation}
\alpha_{{\mathbf s}_i}({\mathbf g})=\int_{|{\mathbf r}-{\mathbf
s}_i|<r_s} \rho({\mathbf r}) e^{-i{\mathbf g}.{\mathbf r}}d^3{\mathbf
r}
\label{eq3}
\end{equation}
where $r_s$ is the radius of the reconstruction sphere and
$\rho({\mathbf r})$ is the appropriate charge density, reconstructed
or pseudo.  The original pseudo-charge densities are available in
reciprocal space directly from the plane-wave calculation, and the
reconstructed charge densities are given as an expansion in spherical
harmonics \cite{trail99}, which allows Eq.\ (\ref{eq2}) and
(\ref{eq3}) to be evaluated.

For an atom situated at the origin Eq.\ (\ref{eq3}) takes the form
\begin{equation}
\alpha_{{\mathbf 0}}({\mathbf g})
=4\pi\sum_{L}(-i)^{l}Y_{L}(\hat{{\mathbf
g}})\int_{0}^{r_s}\rho_{L}(r)j_{l}(gr)r^2dr
\label{eq4}
\end{equation}
where the charge density has been explicitly written as an expansion
in spherical harmonics, and the identity
\begin{equation}
e^{i{\mathbf q}.{\mathbf r}}= 4\pi \sum_{L} i^l j_l(qr)
Y_{L}^*(\hat{\mathbf q}) Y_{L}(\hat{{\mathbf r}})
\end{equation}
has been used.  The radial integral in Eq.\ (\ref{eq4}) is carried out
numerically.  In our calculations for silicon we choose the primitive
unit cell, and reconstruct the core region of one atom chosen to be at
the origin, hence the integral in Eq.\ (\ref{eq4}) is carried out over
a sphere centred on this atom.  Other atoms within the unit cell must
also be taken into account, and in the case of silicon there is
another atom at $\left(\frac{1}{4},\frac{1}{4},\frac{1}{4}\right)$
related to the origin by an inversion symmetry at
$\left(\frac{1}{8},\frac{1}{8},\frac{1}{8}\right)$.  The contribution
to Eq.\ (\ref{eq2}) from this atom can be derived from the symmetry of
the unit cell.  If the atom at the origin is related to an atom at
site ${\mathbf s}$ by the space group operator $\{P|{\mathbf s}\}$
(defined by $\{P|{\mathbf s}\}f({\mathbf r})=f(P{\mathbf r}+{\mathbf
s})$) \cite{altmann91}, where $P$ is a unitary transformation, then
the integral, $\alpha_{{\mathbf s}}$ is
\begin{equation}
\alpha_{{\mathbf s}}({\mathbf g})=\int_{|\{P|{\mathbf
s}\}^{-1}{\mathbf r}|<r_s} e^{-i{\mathbf g}.{\mathbf r}} \{P|{\mathbf
s}\}^{-1}\rho({\mathbf r})\ d^3{\mathbf r}.
\end{equation}
By transforming coordinates this reduces to
\begin{equation}
\alpha_{{\mathbf s}}({\mathbf g})= \int_{|{\mathbf r}|<r_s}
\rho(P^{-1}{\mathbf r}) e^{-i{\mathbf g}.({\mathbf r}+{\mathbf s})}
d^3{\mathbf r}.
\end{equation}
For silicon the atom at
$\left(\frac{1}{4},\frac{1}{4},\frac{1}{4}\right)$ is related to the
atom at the origin by the operator
$\left\{-I|\left(\frac{1}{4},\frac{1}{4},\frac{1}{4}\right)\right\}$,
an inversion followed by a translation.  In this case the above
expression, together with the expansion around the origin in spherical
harmonics, yields
\begin{equation}
\alpha_{{\mathbf s}} ({\mathbf g}) =4\pi e^{-i{\mathbf g}.{\mathbf s}}
\sum_{L}(-1)^{l}(-i)^{l}Y_{L}(\hat{{\mathbf g}})
\int_{0}^{r_s}\rho_{L}(r)j_{l}(gr)r^2dr
\label{eq5}
\end{equation}
where ${\mathbf s}=\left(\frac{1}{4},\frac{1}{4},\frac{1}{4}\right)$.
The transformation results in the phase factor, and the inversion
results in the power of $(-1)$ present in the sum.

Eq.\ (\ref{eq4}) and (\ref{eq5}) are applied to both the reconstructed
charge density and the pseudo-density (expanded in spherical
harmonics) and are then substituted into Eq.\ (\ref{eq2}) to yield the
structure factor as a function of the reciprocal lattice vector,
${\mathbf g}$.  At first it seems a roundabout route to calculate the
radial expansion of the pseudo-density only to convert this back to a
reciprocal space representation, but this is the most straightforward
way of replacing the pseudo-charge density with the reconstructed
charge density in the sphere around each atom.  One final point is the
position of the origin.  The coordinate system used for the
reconstruction has the origin on one of the silicon atoms in the unit
cell (at $\bar43m$), whereas the system normally chosen for
crystallographic studies has the origin at the inversion centre,
($\bar3m$) \cite{hahn95}.  Placing the origin at the inversion centre
gives real structure factors, and the origin can easily be shifted to
this point by introducing an appropriate phase factor into Eq.\
(\ref{eq1}), or simply by taking the magnitude of the complex
structure factors.

\subsection{Comparison with Experimental Results and FLAPW Calculations}
Before comparison can be made between the theoretical and experimental
results two further factors must be considered.  First, the
experimentally measured quantity (normally given in the literature) is
not the Fourier coefficient of the charge density, $\rho({\mathbf
g})$, but the form factor $f_{hkl}$, which takes into account the
lattice structure.  This is defined as \cite{zuo97}
\begin{equation}
f_{hkl}=\rho_{total}({\mathbf g})/ \cos\left(
\left(h+k+l\right)\frac{\pi}{4} \right)
\end{equation}
where $(hkl)$ are the indices of the reciprocal lattice vector.  For
$(hkl)$ values that satisfy the criteria $h+k+l=4n+2$ for $n$ integer,
the denominator on the right hand side is zero, and the structure
factor is given.

The second effect that must be taken into account when correlating the
theoretical and experimental results is the thermal motion of the
lattice.  The majority of experimental data for structure factors are
taken at room temperature, and the thermal energy `smears out' the
charge density, reducing the amplitude of the higher order structure
factors.  This can be described by a convolution integral in real
space, which corresponds to a further correction factor in reciprocal
space to give the \emph{dynamic structure factor}
\begin{equation}
f_{hkl}^{dyn}=f_{hkl} e^{-B g^2/16\pi^2}
\label{eq6}
\end{equation}
where $B$ is the Debye-Waller parameter \cite{deutsch92,lu93,zuo97}.

Structure factors obtained from the core reconstruction are here
compared with those obtained from three sources: from the simple
addition of free atom core states to the original pseudo-charge
density \cite{lu93}, structure factors obtained using the FLAPW method
by Zuo et al \cite{zuo97} and structure factors determined
experimentally by Cumming and Hart \cite{cumming88} and Saka and Kato
\cite{saka86}, as given by Zuo et al \cite{zuo97}.  The pseudo+core
structure factors are obtained from the charge density of the original
pseudopotential calculation together with the core charge densities of
the original atomic calculations used to create the pseudopotential.
The contribution from the atomic core charge density is included at
the atomic sites in the same manner as described above, ie
\begin{equation}
\rho_{total}({\mathbf g})=\rho_{pseudo}({\mathbf g})+\frac{1}{\Omega}
\sum_{i} \alpha^{core}_{{\mathbf s}_i}({\mathbf g})
\end{equation}
where $\alpha^{core}_{{\mathbf s}_i}$ is the contribution from the
core states at site ${\mathbf s}_i$.  This structure factor is
expected to show significant error, since the valence charge density
will be incorrect close to the atomic sites.

Zuo et al \cite{zuo97} calculated Si structure factors using the FLAPW
method, and they give results using the LDA, and two different GGAs.
Since the core reconstruction calculation carried out here employs the
LDA, the reconstruction results are only compared with the LDA FLAPW
results.  For a successful reconstruction scheme we would expect to
accurately reproduce these results, since the same physical
approximations have been made even though the algorithmic
implementation of the two methods is entirely different.

\begin{figure}
\begin{center}
\includegraphics*{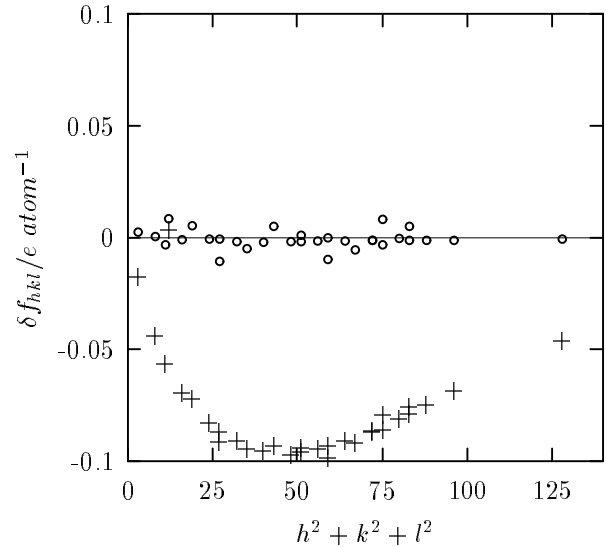}
\end{center}
\caption{Difference between static form factors calculated by
reconstruction and FLAPW methods (circles), and the difference between
static form factors calculated by pseudo+core and FLAPW methods
(crosses).}
\label{fig1}
\end{figure}

We begin by comparing the theoretical form factors, uncorrected for
temperature, and only for the $(hkl)$ values for which experimental
data are available (experimental data are given in Table \ref{tab1}).
Fig. \ref{fig1} shows the difference between the reconstructed and
FLAPW form factors, and the difference between the pseudo+core and
FLAPW form factors.  It is apparent that the reconstruction agrees
very well with the FLAPW results - the average absolute difference for
the reconstructed results is only 3 milli-electrons/atom whereas for
the pseudo+core result the average absolute difference is over 25
times greater at 76 milli-electrons/atom.

\subsection{Experimental, FLAPW and Reconstructed Structure Factors}
In order to compare the static structure factors given above with
experimental data a value for the Debye-Waller parameter in Eq.\
(\ref{eq6}) is required.  This is commonly taken to be a free
parameter and varied to minimise the error between the experimental
and theoretical results.  The value of $B$ used here is that employed
by Zuo et al\cite{zuo97}, \mbox{ $B=0.4668\ $\AA$^2$ }.  In their
paper values of $B$ are obtained by minimising the error of high
$|{\mathbf g}|$ values only, for a number of different \emph{ab
initio} methods.  These high $|{\mathbf g}|$ structure factors depend
largely on the core states of the atoms that make up the lattice, so
the best values should result from methods that accurately describe
the core states.  Zuo et al found that a calculation of these high
order structure factors using the Multi Configuration Dirac-Fock
(MCDF) method \cite{grant80} gives the best fit at high $|{\mathbf
g}|$, and therefore took the associated $B$ parameter to be the best
estimate.  It should be noted that a better fit between experiment and
theory can be obtained for a different $B$ value, but this would
effectively use the description of a physical effect, the thermal
smearing, to adjust for deficiencies in the theory, such as the LDA.

Table \ref{tab1} gives the experimental data, with reconstructed,
FLAPW and pseudo+core dynamic form factors.  The quality of the
theoretical data is assessed by two statistics - the $R$-factor and
GOF parameter.  The $R$-factor is given by
\begin{equation}
R=\frac{\sum_i|f^{theory}_i-f^{exp}_i| } {\sum_i |f^{exp}_i| }
\end{equation}
and the goodness of fit parameter by
\begin{equation}
GOF=\frac{1}{N}\sum_{i=1}^{N} (1/\sigma_i^2)(f^{theory}_i-f^{exp}_i)^2
\end{equation}
where $\sigma_i^2$ is the sample variance of the $i^{th}$ form factor.
The variance $\sigma_i^2$ is taken to be the average of the estimated
error for all data points in line with the approach of Zuo et al, and
takes the value $(0.0022)^2/ e^2\textrm{ }atom^{-2}$.

\begin{table}
\begin{tabular}{crrrr} \hline \hline
           &\multicolumn{4}{c}{$f_{hkl}^{dyn}/ e\textrm{ }atom^{-1}$}
\\ (hkl) & Experimental & Reconstructed & FLAPW & Pseudo+core \\
\hline 1 1 1 & 10.6025(29) & 10.6020 & 10.5995 & 10.5824 \\ 2 2 0 &
8.3881(22) & 8.3955 & 8.3952 & 8.3531 \\ 3 1 1 & 7.6814(19) & 7.6879 &
7.6909 & 7.6373 \\ 2 2 2 & 0.1820(10) & 0.1695 & 0.1615 & 0.1650 \\ 4
0 0 & 6.9958(12) & 6.9924 & 6.9933 & 6.9287 \\ 3 3 1 & 6.7264(20) &
6.7081 & 6.7031 & 6.6365 \\ 4 2 2 & 6.1123(22) & 6.0890 & 6.0897 &
6.0145 \\ 3 3 3 & 5.7806(21) & 5.7456 & 5.7552 & 5.6732 \\ 5 1 1 &
5.7906(27) & 5.7754 & 5.7761 & 5.6984 \\ 4 4 0 & 5.3324(20) & 5.3119 &
5.3136 & 5.2339 \\ 5 3 1 & 5.0655(17) & 5.0447 & 5.0490 & 4.9670 \\ 6
2 0 & 4.6707(9) & 4.6542 & 4.6561 & 4.5748 \\ 5 3 3 & 4.4552(11) &
4.4485 & 4.4444 & 4.3661 \\ 4 4 4 & 4.1239(18) & 4.1069 & 4.1085 &
4.0285 \\ 7 1 1 & 3.9282(22) & 3.9213 & 3.9229 & 3.8449 \\ 5 5 1 &
3.9349(34) & 3.9255 & 3.9248 & 3.8482 \\ 6 4 2 & 3.6558(54) & 3.6413 &
3.6427 & 3.5671 \\ 7 3 1 & 3.4919(11) & 3.4868 & 3.4869 & 3.4135 \\ 5
5 3 & 3.5055(14) & 3.4805 & 3.4883 & 3.4108 \\ 8 0 0 & 3.2485(34) &
3.2458 & 3.2470 & 3.1766 \\ 7 3 3 & 3.1270(14) & 3.1112 & 3.1154 &
3.0453 \\ 8 2 2 & 2.9111(15) & 2.9096 & 2.9105 & 2.8456 \\ 6 6 0 &
2.9143(16) & 2.9095 & 2.9105 & 2.8458 \\ 5 5 5 & 2.8009(21) & 2.8008 &
2.7947 & 2.7361 \\ 7 5 1 & 2.8006(25) & 2.7951 & 2.7976 & 2.7341 \\ 8
4 0 & 2.6200(7) & 2.6216 & 2.6219 & 2.5631 \\ 9 1 1 & 2.5325(8) &
2.5232 & 2.5242 & 2.4678 \\ 7 5 3 & 2.5274(29) & 2.5264 & 2.5229 &
2.4688 \\ 6 6 4 & 2.3677(9) & 2.3724 & 2.3733 & 2.3208 \\ 8 4 4 &
2.1506(24) & 2.1572 & 2.1581 & 2.1115 \\ 8 8 0 & 1.5325(26) & 1.5365 &
1.5370 & 1.5095 \\ \hline R/\% & - & 0.24 & 0.24 & 1.66 \\ GOF & - &
37 & 31 & 1158 \\
\end{tabular}
\caption{Dynamic form factors from experiment, reconstruction, FLAPW
and Pseudo+core calculations.  Estimated errors of the experimental
data are given in parentheses, and the experimental and FLAPW data are
taken from Zuo et al \cite{zuo97}.  A Debye-Waller parameter of \mbox{
$B=0.4668\ $\AA$^2$ } is used, as calculated by Zuo et al
\cite{zuo97}.}
\label{tab1}
\end{table}

\begin{figure*}
\begin{center}
\includegraphics*{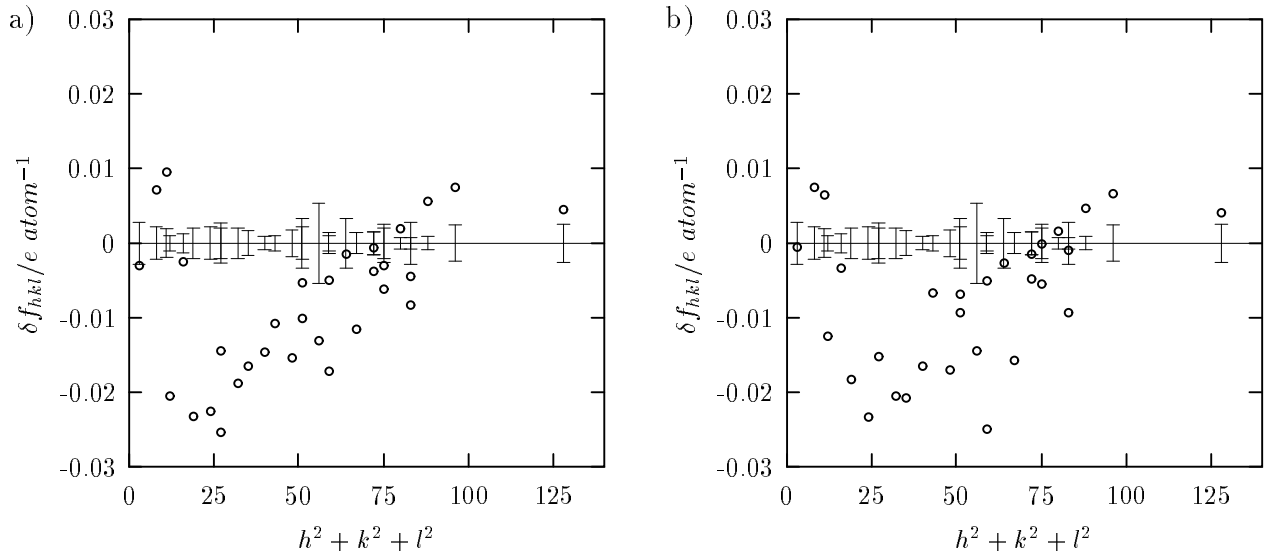}
\end{center}
\caption{Residual error ($f^{theory}-f^{exp}$) of (a) FLAPW and (b)
reconstructed dynamic form factors, with error bars of experimental
data shown.  Note the change of scale from Fig. \ref{fig1}.}
\label{fig2}
\end{figure*}

From the data in Table \ref{tab1} it can be seen that the
reconstruction calculation describes the experimental data as well as
the FLAPW results.  For both sets of data the $R$-factor is
\mbox{$0.24$ \%}, and the GOF is \mbox{$\sim 35$}, with the GOF for
the reconstruction slightly greater than that for FLAPW.  The average
absolute error $|f^{theory}-f^{exp}|$ is 10 milli-electrons/atom for
both the FLAPW and reconstruction calculations and 70
milli-electrons/atom for the pseudo+core results.  The maximum error
is roughly $\sim 20$ milli-electrons/atom for the FLAPW and
reconstruction results, and $\sim 100$ milli-electrons/atom for the
pseudo+core results.  Fig.\ \ref{fig2}a shows the residual error
($\delta f=f^{theory}-f^{exp}$) of the FLAPW results together with the
error bars of the experimental data, and Fig.\ \ref{fig2}b the
residual error for the reconstruction calculation.  The errors are
very similar, even to the point of a significant correlation existing
between the two.  This suggests that the errors present are largely
due to the theory shared by the calculations, specifically the LDA.
It should also be noted that the data presented by Zuo et al is
calculated for a lattice constant of \mbox{$a_0=5.4307$ \AA}, whereas
the reconstruction calculations are carried out for \mbox{$a_0=5.4300$
\AA}.

Finally we give the $R$-factor and GOF parameter comparing the
pseudo+core and reconstructed results with the FLAPW results.  The
pseudo+core form factors give a $R$-factor and GOF of \mbox{$1.55$ \%}
and $1349$ respectively, while the reconstruction gives \mbox{$0.06$
\%} and $3.6$.

\subsection{Spherical Symmetry}
\begin{figure}
\begin{center}
\includegraphics*{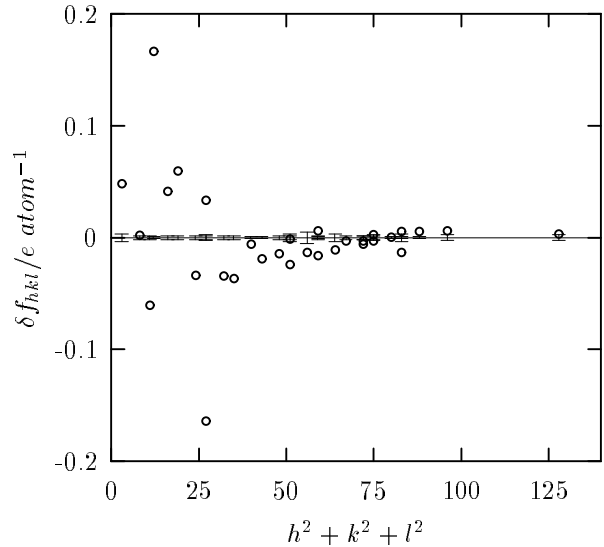}
\end{center}
\caption{Residual error ($f^{theory}-f^{exp}$) of reconstructed
dynamic form factors resulting from updating the pseudo-density with
the \emph{spherical} part of the reconstructed charge density.  Note
that for low $|{\mathbf g}|$ the results are worse than for the
pseudo+core results in Fig. \ref{fig1}.}
\label{fig3}
\end{figure}

One of the strengths of our reconstruction method is that it does not
require spherical symmetry of the charge density in the reconstruction
region near the cores of the atom.  To assess the importance of the
aspherical components of the charge density we replace the original
pseudo-density in the reconstruction sphere with the spherical part
only of the reconstructed density.  Fig.\ \ref{fig3} gives the
residual error of the reconstructed form factors from the experimental
data for this case.  The $R$-factor is $0.64$ \% with a GOF of $485$ -
considerably worse than for the full aspherical reconstruction.  From
this data it is apparent that the aspherical components of the charge
density are essential for the calculation of accurate form factors.
However, it is interesting to note that if we replace the
\emph{spherical} part of the pseudo-density with the \emph{spherical}
part of the reconstructed density (that is the charge density
components $\rho_L$ for $l>0$ within the embedding sphere are given by
the \emph{pseudo}-charge density) we obtain form factors that are
almost as accurate as the FLAPW and fully aspherical results.  In this
case the $R$-factor is $0.25$ \%, the GOF is $37$ and the mean
absolute error is $\sim 11$ milli-electrons/atom).

\section{Conclusions}
In this paper all-electron states have been reconstructed successfully
from a total energy pseudopotential calculation, giving an accurate
charge density in the region near atomic sites.  This reconstruction
is carried out using the embedding method described in a recent paper
\cite{trail99}.  The reconstruction calculation itself uses a scalar
relativistic description for the valence states, in a fully aspherical
potential and using LAPW basis functions.  The core states are
calculated fully relativistically by direct solution of the Dirac
equation in a spherical average of the self consistent potential.  It
is apparent that the reconstruction method itself has a lot in common
with FLAPW methods.

Structure factors have been derived from the reconstructed silicon
charge density for comparison with accurate experimental data and
FLAPW calculations.  Agreement is excellent with both the FLAPW and
reconstructed form factors agreeing with experimental results with an
average absolute error of $10$ milli-electrons/atom while the
experimental data itself is accurate to $3-5$ milli-electrons/atom.
The FLAPW and reconstructed form factors agree extremely well with
each other, with an average absolute difference of $3$
milli-electrons/atom.  In addition to this the residual errors for
both methods of calculation show significant correlation, indicating
that they arise from the physical approximations common to both
methods.

\begin{acknowledgements}
This work has been supported by the United Kingdom Engineering and
Physical Sciences Research Council.  We thank S. Crampin and
J. E. Inglesfield for helpful discussions.
\end{acknowledgements}


\end{document}